# Surprises from Quarkonium Decay into Photons[*]

S. Catani[a] and F. Hautmann[b]

[a]INFN, Sezione di Firenze, Largo E. Fermi 2, I-50125 Florence, Italy

[b]Cavendish Laboratory, University of Cambridge, Madingley Road, Cambridge CB3 0HE, UK

The perturbative QCD approach to quarkonium decay into a photon and hadrons is reconsidered. It is shown that a consistent treatment within perturbative QCD calls for the introduction of a fragmentation contribution which has been neglected so far. The ensuing phenomenological implications are discussed, and, in particular, the possibility of measuring the gluon fragmentation function of the photon is addressed.

## 1. Introduction

One of the earliest applications of perturbative QCD was the calculation of the decay widths of heavy-quark-antiquark bound states (quarkonia) [1]. The use of perturbative QCD in this context is based on the fact that these decay processes involve two very different physical scales: the binding energy $\varepsilon$ of the bound state and the heavy-quark mass $m_Q$. Since $m_Q \gg \varepsilon$ for the $\Psi$- and $\Upsilon$-system, their decay width $\Gamma$ into a given final state $f$ can be factorized as follows

$$\Gamma(X_{Q\bar{Q}} \to f) = R_0^2 \, |M_{Q\bar{Q} \to f}|^2 + \mathcal{O}(v^2/c^2) \; . \quad (1)$$

Here $R_0$ is the wave function of the bound state at the origin, whilst $M$ is the matrix element describing the annihilation of the on-shell heavy quark and antiquark into the final state $f$.

As long as the relativistic corrections $\mathcal{O}(v^2/c^2)$ in (1) are small, all of the bound-state effects are factorized into the wave function, and cancel if one considers ratios of decay widths for different final states. In this case one is thus left with the evaluation of the matrix element $M$, which is dominated by large momentum scales (of the order of $m_Q$). This term is computable (modulo hadronization corrections) as a perturbative series in $\alpha_S(m_Q)$ provided it is infrared and collinear safe, that is, the decay process is *fully inclusive* with respect to the final-state partons.

A well-known example of decay rate which is computable perturbatively is the ratio between the hadronic and leptonic widths. It has been evaluated in leading (LO) and next-to-leading (NLO) order in QCD perturbation theory [2,3] and is mainly used for $\alpha_S$-determinations [4–6].

In this paper we concentrate on the inclusive radiative decay

$$\text{quarkonium} \to \gamma + \text{hadrons} \; , \quad (2)$$

where $\gamma$ denotes a prompt photon, i.e. a final-state photon not coming from hadronic (essentially, $\pi$ and $\eta$) decays. According to common wisdom, the decay rate for the process in Eq. (2) is computable in QCD perturbation theory and even a NLO result is available in the literature [7,8]. Despite common wisdom, we would like to point out that *this* perturbative-QCD picture is not correct, at least from the theoretical viewpoint. In fact, it (*surprisingly*) misses out a *leading-twist* and *leading-order* fragmentation component[1].

In the following we clarify this issue and consider its phenomenological implications. A more detailed discussion will appear elsewhere [10].

## 2. $X_{Q\bar{Q}} \to \gamma + \text{hadrons}$ : revised theory

Let us compare the hadronic decay $X_{Q\bar{Q}} \to$ hadrons with the radiative decay in Eq. (2). In the first case the corresponding partonic final state consists of all the possible channels with many gluons and light quark-antiquark pairs and no particular partonic channel is singled out. In the second case we deal with partonic states of the type $\gamma gg$, $\gamma ggg$, $\gamma gq\bar{q}$, etc., where the photon is directly observed in the final state. Since from the viewpoint of perturbative QCD a photon is a parton as much as gluons and quarks, we are not performing an incoherent sum over all the possible final states. The final state we are looking at is not fully inclusive, and the corresponding matrix element $M$ in Eq. (1) is *not* collinear safe.

---

[*]Contribution at QCD94 Conference, Montpellier, July 1994, presented by S. Catani.

[1]As far as we know, the need to include a fragmentation component was first pointed out in Ref. [9].

More precisely, the matrix element $M(Q\bar{Q} \to \gamma + \text{hadrons})$ gets contributions from two different components (Fig. 1). Besides the usual *direct* component in which the photon is radiated from the heavy quark, there is an additional *fragmentation* component in which the photon is emitted by final-state light quarks. The fragmentation component is obviously of leading-twist order (i.e., not suppressed by inverse powers of $m_Q$ in the limit $m_Q \to \infty$) and is not finite in perturbation theory. There are indeed singularities from the integration region where the photon is collinear to the radiating quark. The only consistent way of dealing with these singularities is to *factorize* them into non-perturbative fragmentation functions of the photon $D(z, m_Q)$.

Before presenting the complete formalism which includes the fragmentation function, let us estimate the order of magnitude of the fragmentation component. The associated collinear singularities are hidden in two- (and higher- ) loop Feynman diagrams of the type in Fig. 2. Here the leading phase-space region is that in which the photon-quark angle $\theta_{\gamma q}$ and the quark-antiquark angle $\theta_{q\bar{q}}$ are both small. Correspondingly we obtain a double-logarithmic singularity of the type $(\ln m_Q^2/Q_0^2)^2$, $Q_0$ being a collinear cut-off of the order of some hadronic scale. Multiplying this contribution by the relevant coupling constant factor, we obtain $\alpha \alpha_S^4(m_Q)(\ln m_Q^2/Q_0^2)^2$. Since $\ln m_Q^2/Q_0^2 \sim 1/\alpha_S(m_Q)$, we see that the collinear fragmentation kills two powers of $\alpha_S$ thus leading to a contribution exactly of the same order as the lowest-order direct term, i.e. $\mathcal{O}(\alpha \alpha_S^2)$. We conclude that the fragmentation component may be of the same size as the *leading-order* direct component.

This argument also shows that in the dominant collinear region $(\theta_{\gamma q}, \theta_{q\bar{q}} \to 0)$, the gluon which initiates the fragmentation cascade is near mass-shell. Therefore the leading fragmentation contribution involves $D_{g\gamma}(z, m_Q)$, the gluon fragmentation function of the photon. Actually, the process in Eq. (2) is the *only* physically relevant process in which the gluon fragmentation function $D_{g\gamma}$, and *not* the quark fragmentation function $D_{q\gamma}$, enters in LO. In fact, in the case of prompt photons produced in $e^+e^-$-annihilation only $D_{q\gamma}$ is involved in LO, whilst both $D_{q\gamma}$ and $D_{g\gamma}$ enter the LO calculation of prompt-photon production by hadron collisions.

As a final comment, we should point out that the total width (i.e. the width integrated over the energy fraction $z = E_\gamma/m_Q$ of the photon) for the process (2) is not only collinear unsafe but also *infrared unsafe*. Indeed the fragmentation process has an associated bremsstrahlung spectrum of the type $dz/z$ which cannot safely be integrated down to $z = 0$. Only the photon energy spectrum (away from $z = 0$) is computable in (fixed-order) perturbation theory, provided collinear singularities are factorized into $D_{q\gamma}$ and $D_{g\gamma}$.

## 3. The photon spectrum

The photon spectrum for the decay of the quarkonium state $X_{Q\bar{Q}} = J/\Psi, \Upsilon$ into $\gamma +$ hadrons can be written as follows

$$\frac{d\Gamma}{dz} = \frac{32\,(\pi^2-9)}{9\,\pi} \frac{R_0^2}{M^2} \frac{d\Gamma_\gamma}{dz}(z,M) \,, \qquad (3)$$

$$\frac{d\Gamma_\gamma}{dz}(z,M) = e_Q^2 \alpha\, \alpha_S^2(M)\, \hat{\Gamma}_\gamma\left(z, e_Q; \alpha_S(M)\right) \quad (4)$$

$$+\alpha_S^3(M) \sum_{a=q,\bar{q},g} \int_z^1 \frac{dx}{x} \hat{\Gamma}_a\left(x;\alpha_S(M)\right) D_{a\gamma}(\frac{z}{x}, M) \,.$$

Here $M \simeq 2\,m_Q$ is the quarkonium mass and $e_Q$ is the corresponding heavy-quark charge. The first and the second term on the r.h.s. of Eq. (4) respectively denote the direct and fragmentation contribution[2]. $\hat{\Gamma}_A$ ($A = \gamma, q, \bar{q}, g$) are hard coefficient functions computable as power series expansions in $\alpha_S$:

$$\hat{\Gamma}_A(z;\alpha_S) = \hat{\Gamma}_A^{(0)}(z) + \frac{\alpha_S}{\pi} \hat{\Gamma}_A^{(1)}(z) + \cdots \,. \qquad (5)$$

---

[2]Strictly speaking, the distinction between direct and fragmentation component is factorization-scale dependent. In Eq. (4) we have set both the renormalization and factorization scales equal to $M$.

The direct term to LO is well known [7] and well approximated by a linear spectrum $\hat{\Gamma}_\gamma^{(0)}(z) \simeq 2\,z$. As regards the fragmentation terms $\hat{\Gamma}_a^{(0)}$, we know from Sect. 2 that only $\hat{\Gamma}_g^{(0)}$ is non-vanishing. Its actual computation gives $\hat{\Gamma}_g^{(0)}(z) = 5\,\hat{\Gamma}_\gamma^{(0)}(z)/12$.

None of the NLO coefficients $\hat{\Gamma}_A^{(1)}$ is known at present. The calculation in Ref. [8] refers to the integral $\int_0^1 dz\,\hat{\Gamma}_\gamma^{(1)}(z)$. As for the shape of the direct term $\hat{\Gamma}_\gamma(z)$ in higher orders, only some model calculations are available [11,12]. For these reasons in the following we limit ourselves to considering the photon spectrum in LO.

## 4. Leading-order results

In order to compute the spectrum in Eq. (4) we need the fragmentation functions $D_{a\gamma}(z,M)$ of the photon. Perturbative QCD predicts only their $M$-dependence via evolution equations of Altarelli-Parisi type [13]. The solution of these equations depends on non-perturbative fragmentation functions at some input scale $Q_0^2$. The latter are measurable in principle but unknown at present. In our analysis we use a reasonable set of LO[3] fragmentation functions as provided by Owens [13]. This set is obtained by choosing the input $D_{a\gamma}(z,Q_0 = \Lambda) = 0$, $\Lambda$ being the QCD scale, and is parametrized as follows [13]

$$z\,D_{a\gamma}(z,M) = \frac{6}{33 - 2\,N_f}\,\frac{\alpha}{\alpha_S(M)}\,f_a(z)\ . \qquad (6)$$

Note that the functions $f_a(z)$ are scale invariant, so that the full $M$-dependence is due to the factor in front of them on the r.h.s. of Eq. (6). The actual form of $f_a(z)$ given in [13] shows that the gluon (quark) fragmentation dominates at small (large) values of $z$.

The LO photon spectra obtained by inserting Eq. (6) into Eq. (4) are presented in Fig. 3. Note that, because of the dependence $D_{g\gamma} \propto \alpha/\alpha_S$, the ratio between direct and fragmentation contribution is independent of $\alpha_S(M)$ (the value of $\alpha_S(M)$ only affects the overall normalization in Fig. 3). Note also that this ratio is proportional to $e_Q^2$ (see Eq. (4)), so that the relative effect of the fragmentation is larger for $\Upsilon$ than for $J/\psi$.

---
[3]NLO fragmentation functions are also available [14], but they are not relevant in our LO analysis.

The modification of the LO direct spectrum due to the fragmentation component reflects the $z$-behaviour of the gluon fragmentation function. The latter is very soft and, hence, the fragmentation contribution is dominant at small $z$ but quite small at large $z$. On the other hand, accurate experimental data [15,16] are available at present only in the large-$z$ region. It follows that the LO fragmentation term considered here does not have a big impact on the QCD studies performed so far (typically, only the data points for $z \gtrsim 0.5$ have been used in these analyses).

The same conclusion, however, does not necessarily apply beyond LO. In fact, although the NLO fragmentation term is still unknown, also the quark fragmentation function of the photon is involved in this order. According to the Owens parametrization [13], $D_{q\gamma}$ is likely to be two orders of magnitude larger than $D_{g\gamma}$ at $z \gtrsim 0.6$. Therefore the NLO fragmentation component is not necessarily small in the large-$z$ region. Moreover, the NLO direct contribution to $d\Gamma/dz$ is also unknown and, more importantly, one cannot concentrate on very large values of $z$ in the attempt of suppressing the effect of fragmentation. In the 'extreme'-$z$ region (say, $1 - z \lesssim (1\,\mathrm{GeV})/M$, i.e., $z \gtrsim 0.7$ for $J/\psi$ and $z \gtrsim 0.9$ for $\Upsilon$), hadronization effects are much larger than contributions computable in perturbation theory.

As a result of this discussion, we regard the present tests of perturbative QCD (in NLO) from the process (2) as quite uncertain.

## 5. Conclusions

We have shown that the perturbative QCD approach to the radiative decay of quarkonia requires the introduction of a fragmentation component (surprisingly never considered so far) besides the usual direct contribution. Such fragmentation component enters to leading-twist order and to LO in $\alpha_S$. The perturbative QCD predictions for the photon spectrum thus depend on $\alpha_S$ and the fragmentation functions $D_{a\gamma}$ of the photon. In principle, measuring the photon spectrum at a single scale ($J/\psi$ or $\Upsilon$ decay), one cannot extract a value for $\alpha_S$. Measurements from both $J/\psi$ and $\Upsilon$ decays, in contrast, allow a combined determination of $\alpha_S$ and $D_{a\gamma}$. In practice, the fragmentation contribution is small at large $z$, so that data in this region can be used for $\alpha_S$-determinations.

On the other hand, fragmentation dominates at small values of $z$. In this region it is experimentally very difficult to disentangle the prompt-photon component from the large background due to $\pi$ and $\eta$ decays. Nevertheless, further experimental and phenomenological investigations of this issue can be extremely interesting. Data at small $z$ could indeed be used for measuring (or setting bounds on) the *gluon* fragmentation function $D_{g\gamma}$ of the photon. The perturbative QCD predictions for prompt-photon cross sections in hadron collisions substantially depend on $D_{g\gamma}$, and at present we do not have any experimental information on it.

As for $\alpha_S$-determinations from quarkonium decay, we would like to add a last comment. The ratio between the hadronic and leptonic widths is known up to NLO in perturbative QCD but, due to the different topologies of the hadronic- and leptonic-decay mechanisms, it is likely to be affected by large relativistic corrections [4,5,17]. These corrections are expected to be smaller for the ratio between the radiative and hadronic widths, but, as discussed in this paper, a complete NLO calculation for the photon spectrum is still missing. For these reasons we recommend (see, for instance, Refs. [5,18]) a careful and 'generous' estimate of the theoretical uncertainties on the values of $\alpha_S$ extracted from quarkonium decays.

**Acknowledgements.** Valuable discussions with M. Consoli, M. Kobel and B.R. Webber are gratefully acknowledged. We wish to thank Stephan Narison for the pleasant and stimulating atmosphere at this Conference. This research is supported in part by the EEC Programme *Human Capital and Mobility*, Network *Physics at High Energy Colliders*, contract CHRX-CT93-0357 (DG 12 COMA).

Fig. 1. (a) Direct and (b) fragmentation contributions to the matrix element $M(Q\bar{Q} \to \gamma +$ hadrons).

Fig. 2. The lowest-order Feynman diagram leading to photon fragmentation.

Fig. 3. LO results (in units $R_0/M = 1$: see Eqs. (3) and (4) ) for the prompt-photon spectrum in (a) $J/\psi$ and (b) $\Upsilon$ decays, with $\alpha_S = 0.25$ and $\alpha_S = 0.20$ respectively.

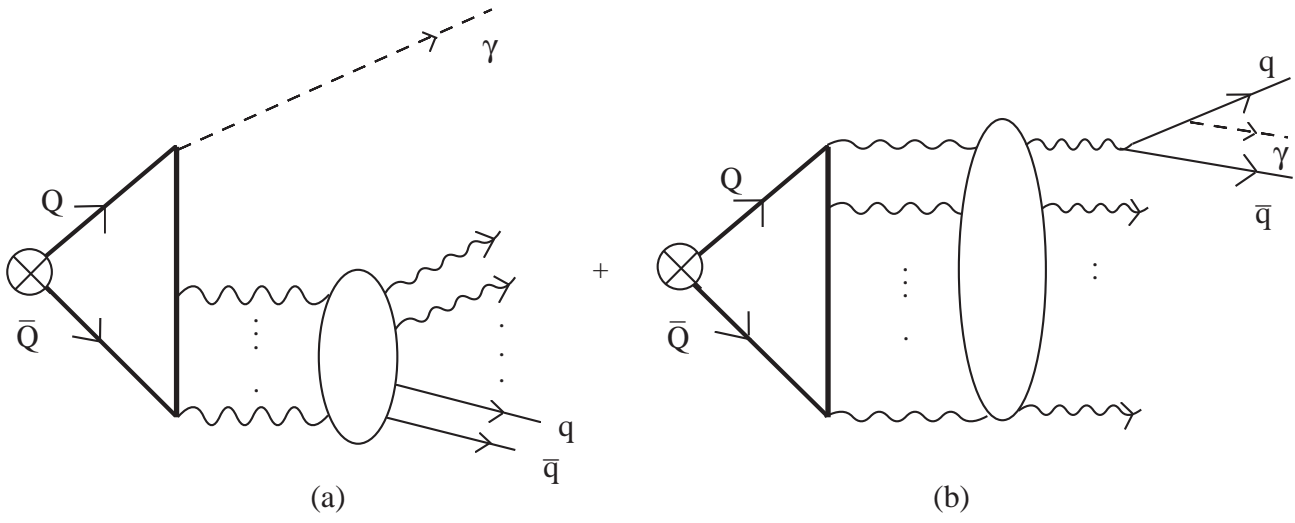

Fig. 1

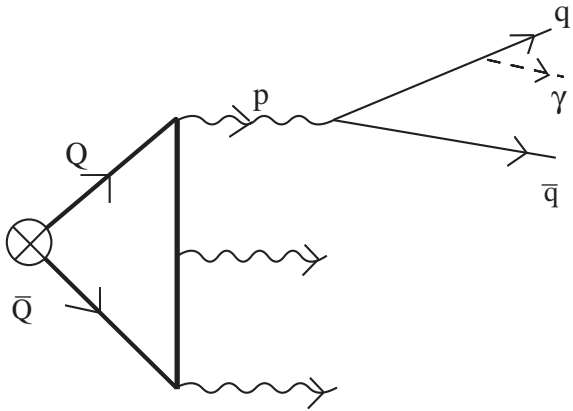

Fig. 2

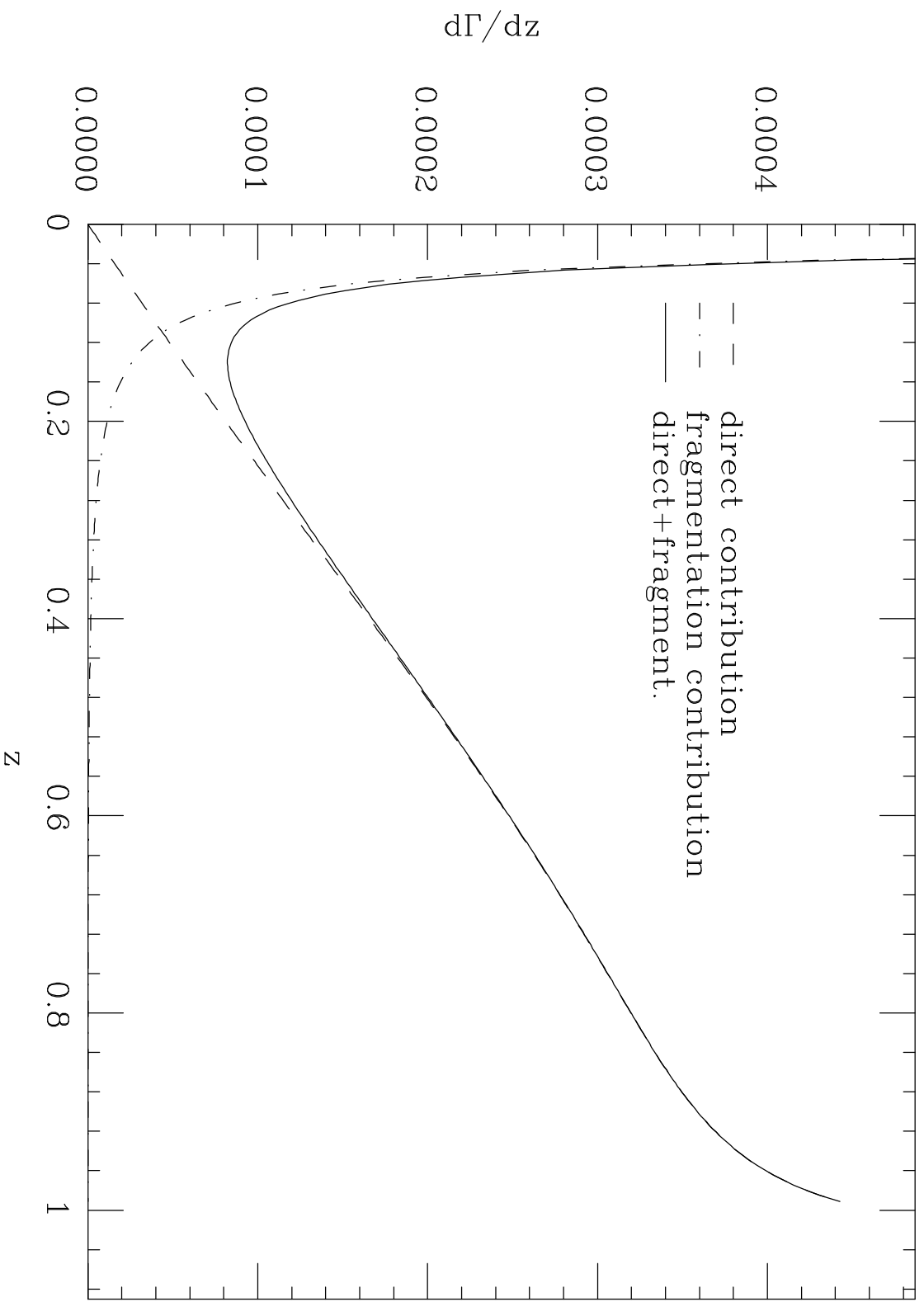

Fig.3a

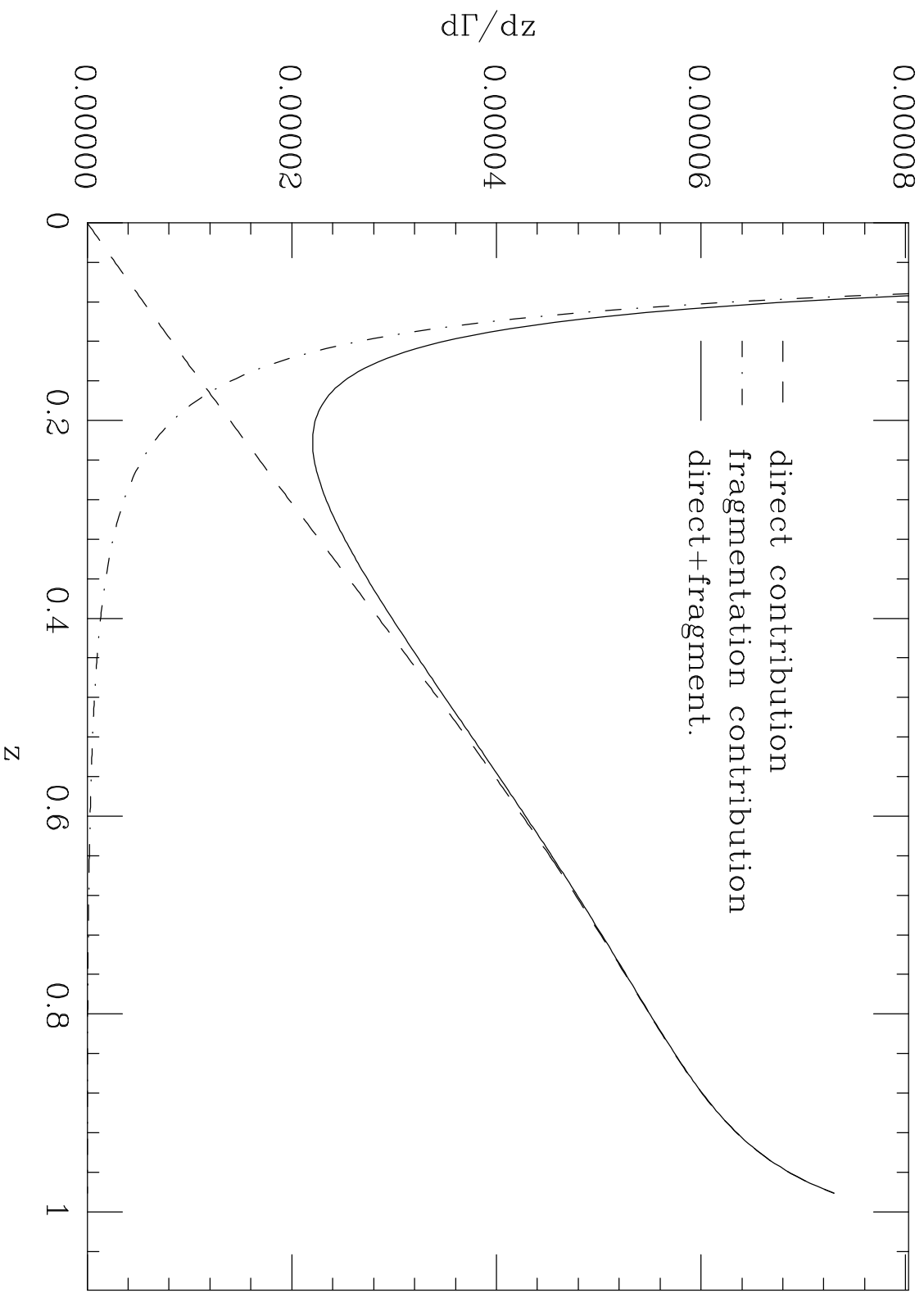

Fig.3b